\documentclass[twocolumn,amsmath,amssymb,prl,superscriptaddress,a4paper,floatfix,showkeys]{revtex4}
\usepackage{graphicx}
\usepackage{color}
\usepackage{mathptmx}      

\begin{document}

\title{On the roles of Vorob'ev cyclicities and Berry's phase in the EPR paradox and Bell-tests}

\author{David H. Oaknin}
\email{d1306av@gmail.com}
\affiliation{%
Rafael Ltd., IL-31021 Haifa, Israel
}%
\author{Karl Hess}
\email{k-hess@illinois.edu}           
\affiliation{%
Center for Advanced Study, University of Illinois, Urbana, Il 61801, USA
}%


\begin{abstract}
The well known inequalities of John S. Bell may be regarded, from a purely mathematical viewpoint, as a direct consequence of Vorob'ev-type topological-combinatorial cyclicities formed with functions on a common probability space. However, the interpretation of these cyclicities becomes more subtle when considerations related to gauge symmetries and geometric-combinatorial phases are taken into account. These physics related considerations permit violations of all Bell-type inequalities within the realm of Einstein's causal physical-mathematics.
\end{abstract}
\maketitle

Many recent books on quantum theory and quantum informatics dedicate considerable sections to the Einstein-Podolsky-Rosen (EPR) \cite{EPR} paradox and Bell's inequalities \cite{Bell, bell}, which appear to represent the ultimate roadblock for an intuitive explanation of quantum phenomena. They seem to form the mathematical embodiment of Bohr's enunciation that the atomic world cannot be explained by using the physical concepts of our macroscopic experiences and the corresponding mathematical language. In turn, removing this roadblock might lead to an interpretation of the quantum formalism without forfeiting the fundamental physical principles that lay behind our views of the macroscopic world. This fact becomes particularly obvious in reports of science writers, who commonly present Bell's inequality as a consequence of straightforward algebra or logic and then struggle with the problem that quantum theory and actual EPRB experiments seem to contradict that logic. Their proposed way out of the conundrum is the introduction of instantaneous influences at a distance (that Einstein called ``spooky") and/or abandoning the notion of physical reality that we ordinarily acknowledge in our macroscopic world.

The non-sequitur of such radical measures has been pointed out in several thoughtful works e.g. in \cite{Scully}.  However, Wigner's \cite{Wigner} set theoretical reasoning appeared difficult to overcome even to these authors and indeed may be overcome only by the detailed mathematical physics given in the bulk of this paper. 

Bell-type inequalities constrain the correlations that may appear in probabilistic models that fulfill certain generic features.  Quantum phenomena appear to violate such constraints and, thus, cannot be described or understood by any of these generic models. This fact is interpreted as an experimentally verifiable proof that quantum phenomena - and, in particular, quantum entanglement - cannot be understood in terms of Kolmogorov's classical probability concepts and fundamental physical principles such as Einstein's notion of causality.   

\section*{Elementary derivation of the key result for Bell tests}

Bell tests involve a number of variations of EPR-type experiments and are thought to present the quintessential demonstrations of whether or not quantum systems may be described in terms of physical concepts taken from the macroscopic world. In such experiments, a source emanates pairs of entangled particles, which propagate toward two distant detection systems that test their polarizations. Each detector may be positioned in one of two available settings defined with respect to local lab frames. Upon detection each detector produces a binary response - either $-1$ or $+1$, so that the correlation between the outcomes at the two measurement stations is given by 
\begin{equation}
\label{the_correlation}
E(\Delta) = -\cos(\Delta),
\end{equation}
where $\Delta$ is the relative angle between the orientations of the two detectors. Eq.[\ref{the_correlation}] represents the results of both quantum theory and many experiments. However, this result is considered to be inconsistent and impossible to obtain with Einstein's causality aided by Kolmogorov's probability theory.
\\

This supposed impossibility is very suspicious, because the correlation [\ref{the_correlation}] can be accounted for on the basis of straightforward symmetry arguments and smoothness constraints as well as standard tools of information theory. In principle, we could expect that the correlation between the outcomes of the two detection systems may depend on their separate orientations, $\Delta_A$ and $\Delta_B$, defined with respect to local lab frames. These orientations can be alternatively described in terms of their relative orientation $\Delta = \Delta_B - \Delta_A$ and their global rigid orientation ${\Gamma} = (\Delta_A + \Delta_B)/2$. Because the pair of entangled particles is invariant under rotations, the correlation between the outcomes of the two detectors (polarizers) may (reasonably speaking) only depend on the relative angle $\Delta$ between them and not on their global rigid orientation $\Gamma$,  and the experimental results indeed show so (with great accuracy). The dependence of the correlation on the relative angle $\Delta$ arises naturally from the simple fact that we compare and statistically collect 'equal' and 'not-equal' experimental outcomes when evaluating the EPRB experiments; a procedure that  involves the measurement results in both wings. This procedure is also basic to Wigner's \cite{Wigner} approach. More detailed explanations will be given in the bulk of this paper.

The probabilities for 'equal' and 'not-equal' outcomes at the two detection systems may then be written, without any loss of generality, as:
\begin{eqnarray}
\begin{array}{cccccc}
p(\mbox{'EQUAL'}) \hspace{0.33in} & = & \sin^2(\chi(\Delta)) & \equiv & \ p_1(\chi(\Delta)), \\
p(\mbox{'NOT-EQUAL'}) & = & \cos^2(\chi(\Delta)) & \equiv & \ p_2(\chi(\Delta)),
\end{array} \label{july22n1}
\end{eqnarray}
so that
\begin{equation}
\label{corrX}
E(\Delta) = \sin^2(\chi(\Delta)) - \cos^2(\chi(\Delta)) = -\cos(2\cdot\chi(\Delta)).
\end{equation}
Since the correlation functions must fulfill the symmetry constraints
\begin{eqnarray}
E(\Delta + \pi) = - E(\Delta) = -E(-\Delta),
\end{eqnarray} 
we postulate that the function $\chi(\Delta)$ fulfills 
\begin{eqnarray}
\label{boundaryconstraints1}
\chi(-\Delta) & = & -\chi(\Delta), \\
\label{boundaryconstraints2}
\chi\left(\Delta + \pi \right) & = & \left(\frac{\pi}{2} \pm \chi(\Delta)\right) \ \left[\mbox{mod} \ [0, \pi)\right].
\end{eqnarray} 
We will see in a section below that the plausible linear relation:
\begin{equation}
\label{Shannon}
 \chi(\Delta) = \frac{\Delta}{2},
\end{equation}
may actually be derived from the above symmetry constraints and well known tools of Informatics. Thus, the quantum correlations [\ref{the_correlation}] are obtained as a variation of the Malus law as suggested in \cite{HESS2020}.
\\

It is the purpose of this paper to show that  Eq.[\ref{the_correlation}] neither contradicts Kolmogorov's set theoretic probability theory nor Einstein's causality principle.  The basis of our findings is the following. Bell-type inequalities have actually been known to mathematicians in one form or another since the early work on probability theory by Boole \cite{Boole} and found their most general formulation in the work of Vorob'ev \cite{Vorob}. Thus, it has been known that the constraints demanded by Bell type inequalities are a consequence of  certain cyclicities of concatenated random variables on a Kolmogorov probability space. The violation of the inequalities implies that the joint probabilities associated to the random variables cannot always be defined on a single probability space, unless the cyclicities may be somehow avoided. We shall show below that the cyclicities involved in the Bell and CHSH \cite{CHSH} theorems may be removed by well known yet subtle physics involving gauge symmetry considerations, geometric phases and other factors \cite{Oaknin,Oaknin1,Oaknin2,HESS18}.

\subsection*{Bell-type inequalities and Vorob'ev cyclicities}

Bell introduced function-pairs related to two measurement stations with certain instrument settings that are usually denoted by ${\bf j}, {\bf j'} = {\bf a}, {\bf b}, {\bf c}, {\bf d}$. One function, $A({\bf j}, \lambda)$, describes the measurement outcomes in station 1 and the function $B({\bf j'}, \lambda)$ describes those in station 2. Here, ${\bf j}$ and ${\bf j'}$ are variables representing the instrument settings. The variable $\lambda$ is, according to Bell, corresponding to elements of physical reality and symbolizes the pair of entangled quantum entities that are sent to the respective instruments from a common source. Therefore, $\lambda$ appears as variable in both functions $A, B$ and describes the information shared by the two stations through the pair of entangled electrons or photons. 
\\

The purpose of the function-pairs $A, B$ is to describe  in mathematical form the Einstein-Podolsky-Rosen (EPR) \cite{EPR} Gedanken-experiment and its variations and it was Bell's intention to link the domain and co-domain of his functions to actual, performable, experiments. In many later publications, it became evident that Bell's functions needed to depend on all possible elements of physical reality that  Einstein's theory of relativity has provided, which includes measurements with rigid rods and with clocks  to describe dynamical effects. Thus, the interpretation of Bell's $\lambda$ has a long and checkered history. However, for the following discussions, it is sufficient to regard $\lambda$ as the mathematical symbol that describes the elements of reality emanating from the source, some of them possibly 'hidden' to our current knowledge. Dynamical effects related to these elements of reality may then be ``absorbed" in the method of their evaluation by measurement equipment and gauge considerations.
\\

Furthermore, Bell wished to connect his mathematical considerations to quantum physics and to actual experiments and, therefore, needed to involve probabilities. He thus assumed that some of his variables may be random variables. In Kolmogorov's probability framework these are functions on a probability space. Mathematical work on probability themes often starts with the words: ``given a probability space $\Omega$" and takes it for granted that such a probability space exists. So did Bell and all the authors following his work. They concluded that the joint probabilities predicted by quantum mechanics for Bell experiments cannot be described using a single probability space. 
\\

\begin{figure}
\begin{center}
\includegraphics[width=10cm]{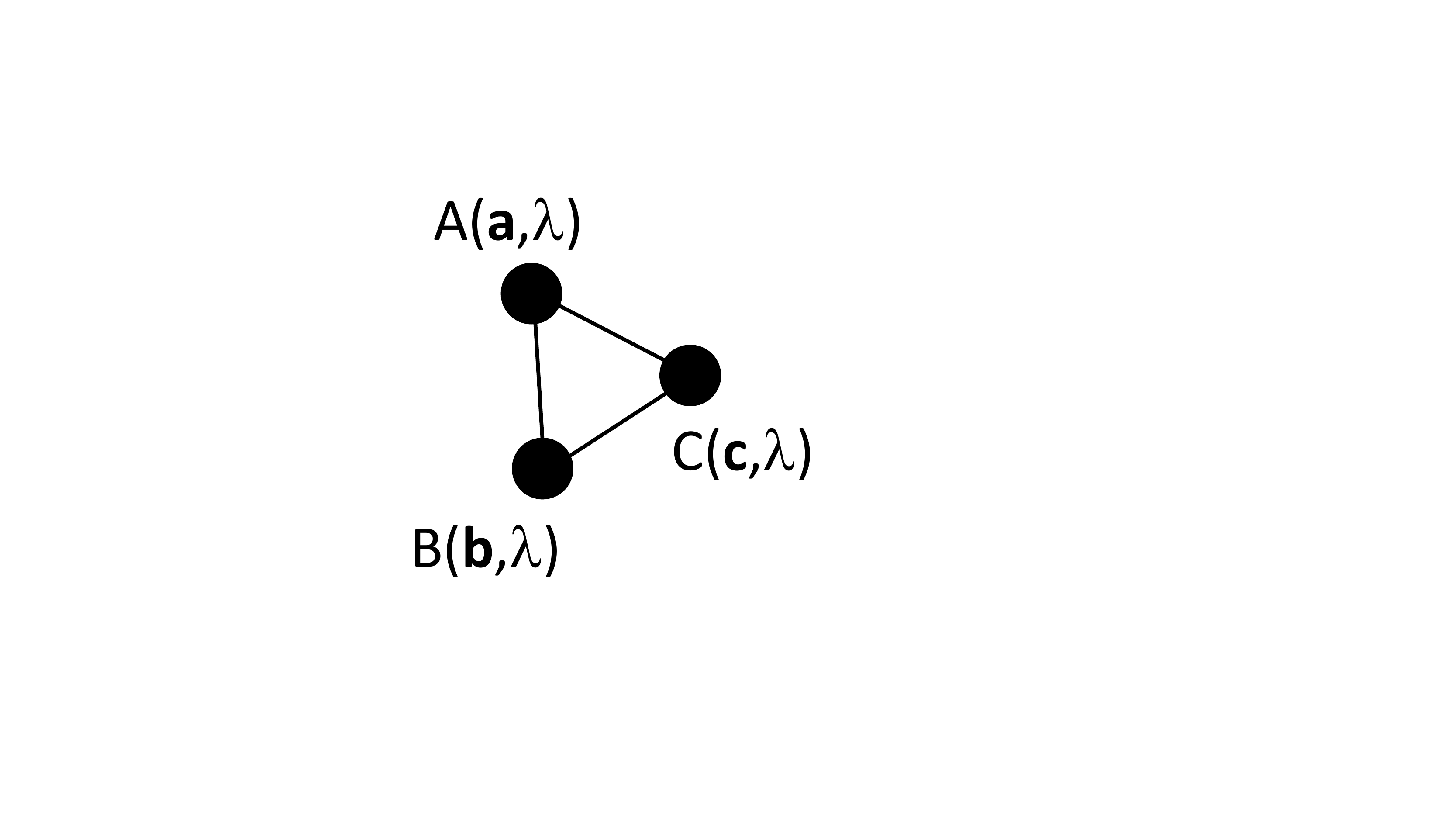}
\end{center}
\caption{Vorob'ev cyclicity for Bell's functions and inequality}
\label{fig:VorobBell}
\end{figure}

In fact, Vorob'ev had shown previously that certain expectation values and corresponding probabilities for function-pair-outcomes cannot consistently exist on a single common probability space $\Omega$, if a combinatorial-topological cyclicity is involved in the concatenation of random variables (functions on a probability space) and they exceed certain constraints. Vorob'ev's generality of argument makes it necessary to involve combinatorial topology. The essence and principle of his reasoning can be made clear from the graphical representation for the special case of Bell's functions shown in Fig.\ref{fig:VorobBell}. The Bell inequality deals then only with the three function-pairs:

\begin{equation}
A({\bf a}, \lambda) \ B({\bf b}, \lambda) \text{;} \ \ A({\bf a}, \lambda) \ C({\bf c}, \lambda) \text{;} \ \ B({\bf b}, \lambda) \ C({\bf c}, \lambda), \label{may13n1}
\end{equation}
Here we have used, with Bell, identical $\lambda$s for all of the three pairs, which is equivalent to the assumption that the cyclically connected functions may be defined on one common probability space. The Vorob'ev cyclicity that corresponds to these three function-pairs is that of the triangle shown in Fig.\ref{fig:VorobBell}. Vorob'ev has emphasized that the arbitrary prescription of joint pair probability-distributions to the first two pairs does not permit complete freedom to choose the joint distribution of the last pair. This fact puts a constraint on the possible pair expectation values in form of Bell-type inequalities. The form of the cyclicity determines the form of the inequalities.
\\

\begin{figure}
\begin{center}
\includegraphics[width=10cm]{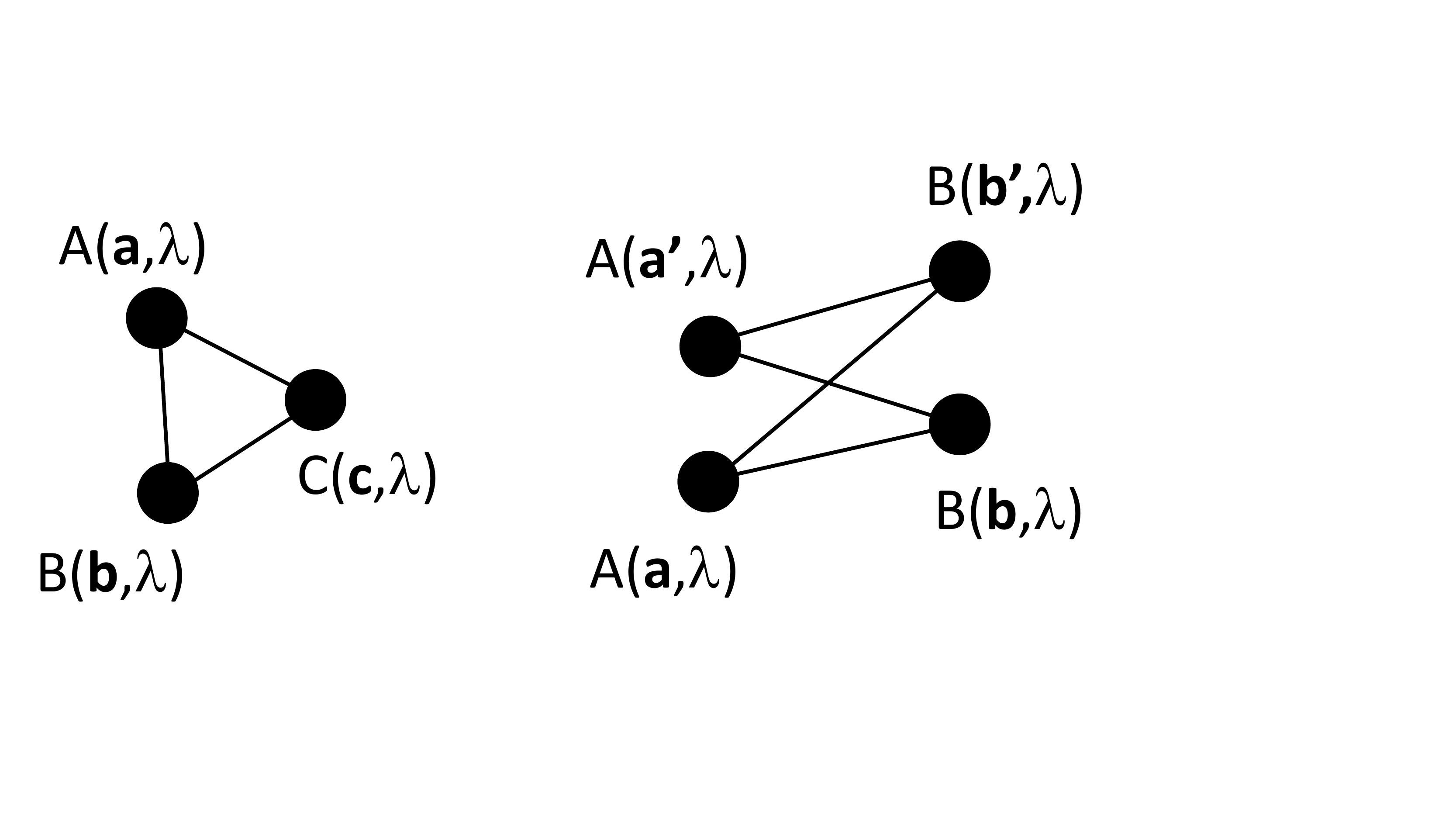}
\end{center}
\caption{Symbolized Vorob'ev cyclicity for both Bell- and CHSH-type inequalities}
\label{fig:VorobCHSH}
\end{figure}

Similar considerations apply to the CHSH inequalities \cite{CHSH} as shown in Fig.\ref{fig:VorobCHSH}. The corresponding cyclicity is represented for four pairs of Bell-type functions:
\begin{equation}
A({\bf a}, \lambda) \ B({\bf b}, \lambda) \text{;} \ \ A({\bf a}, \lambda) \ B({\bf b'}, \lambda) \text{;} \ \ A({\bf a'}, \lambda) \ B({\bf b}, \lambda)
    \label{may14n1}
\end{equation}
and
\begin{equation}
A({\bf a}', \lambda) \ B({\bf b}', \lambda). \nonumber
\end{equation}
The cyclicity imposes again constraints if we restrict ourselves to one common probability space.
\\

Bell's theorem is widely understood as a experimentally testable statement that quantum mechanical joint probabilities for the separate pair-wise measurement outcomes cannot be defined on a single probability space and, hence, there cannot exist an underlying more fundamental description of quantum phenomena. We want to show here that these joint probabilities can be properly defined on a single probability space through the use of the gauge symmetries of the problem. It is our declared purpose to show that Vorob'ev's cyclicities that are inherent in Bell-type constraints can be eliminated by a careful consideration of the involved gauge symmetries, geometric phases and other factors, so that the constraints can be completely avoided.
\\

In order to show how to eliminate the cyclicities it is important to note that in Bell's formulation the variables ${\bf j}, {\bf j'}$ that describe the settings of the instruments, as well as the variables $\lambda$ that describe the elements of reality of the entangled pairs, are defined with respect to local lab frames. However, while the settings ${\bf j}, {\bf j'}$ can be defined with respect to local lab frames, it is not always true that the variables $\lambda$ can be so defined when cyclicities and gauge symmetries are involved. Neither is it necessarily true that the response functions of the instruments can be defined in terms of variables defined with respect to lab frames as assumed in Bell's formulation. In general, the variables $\lambda$ may be properly defined only with respect to the setting of each instrument, and the response of the instrument would then be a function of the variable $\lambda$ defined only with respect to the instrument setting. More detailed explanations will be given below in section {\bf 1.B}. \\

To this regard, we must note that the original EPR argument involves only instruments in parallel settings, for which their outcomes are fully (anti)correlated. Hence, the original EPR argument does not need to independently define the setting of each one of the instruments. Independently defined settings for each one of the instruments were introduced by Bell, not by Einstein and his collaborators. 

\section{Removing the cyclicities and Bell's constraints}

\subsection{Deriving the quantum result: Wigner's counting}

Wigner generalized Bell's procedure by using set theory and let the outcomes include any measurement result, given the instrument settings of both experimental wings. He then used only a judgement of equal or not-equal measurement-outcomes and/or function values (such as spin ``up" versus ``down" or ``horizontal" versus ``vertical", respectively) \cite{Wigner}. As outlined in \cite{HESS2020}, Wigner's counting ensures that only the relative outcomes of the two wings are of importance. This fact also ensures that the physical variable responsible for the correlation between the outcomes of the measurements is the relative angle $\Delta$ between the directions of the polarizers (Stern-Gerlach magnets) and leads to the following equation that was already derived in the introduction:
\begin{equation}
E(\Delta) = \sin^2(\chi(\Delta)) - \cos^2(\chi(\Delta)) = -\cos(2\cdot\chi(\Delta)). \nonumber
\end{equation}
In order to link this equation to the results of quantum theory and experiments, we still have to show the linearity of the function $\chi(\Delta)$. This may be accomplished as follows: Because the Fisher information for the random game discussed in the introduction is given by \cite{Hans, Luo}
\begin{equation}
I_F(\Delta) = \sum_{i=1,2} \frac{1}{p_i(\chi(\Delta))} \cdot \left(\frac{\partial p_i(\chi(\Delta))}{\partial \Delta}\right)^2 = 4 
 \left(\frac{d \chi(\Delta)}{d \Delta}\right)^2,
\end{equation}
it attains a constant value consistent with the symmetry constraints [\ref{boundaryconstraints1},\ref{boundaryconstraints2}] for
\begin{equation}
\chi(\Delta) = \frac {\Delta} {2}.
\end{equation}
In other words, the quantum correlation [\ref{corrX}] corresponds to the situation in which the correlation of the outcomes that are measured in the two stations for the single particle pairs carries the minimum possible information about the relative angle $\Delta$ at which it was obtained.

It is interesting to notice that the symmetry constraints [\ref{boundaryconstraints1},\ref{boundaryconstraints2}] can also be obtained through similar considerations. Because the Shannon entropy for the random game is given by \cite{Shannon1,Shannon2}
\begin{equation}
S[\chi(\Delta)] =-\sum_{i=1,2} p_i(\chi(\Delta)) \cdot \mbox{log}\left(p_i(\chi(\Delta))\right),
\end{equation}
we define the total entropy, with the help of some symmetry considerations, as
\begin{equation}
{\cal Q}[\chi(\Delta)] \equiv \int_{0}^{\pi} d\Delta \ \ S[\chi(\Delta)], 
\end{equation}
whose extrema obey the equation,
\begin{equation}
 \frac{\delta {\cal Q}[\chi(\Delta)]}{\delta \chi(\Delta)} = \int_{0}^{\pi} d\Delta \ \ \frac{\delta S[\chi(\Delta)]}{\delta \chi(\Delta)} = 0.
\end{equation}
The last equation can be written as:
\begin{equation}
\int_{0}^{\pi} d\Delta \ \ \sin(2 \chi(\Delta)) \ \mbox{log} (\left|\tan(\chi(\Delta))\right|) = 0,
\end{equation}
which is fulfilled for all functions $\chi(\Delta)$ that obey the said symmetry constraints.

Similar arguments can be applied to other Bell states beside the singlet state. In all cases there is a true physical angle, which corresponds to a properly defined relative orientation between the measuring devices, while its orthogonal combination is a gauge degree of freedom that corresponds to a global rigid orientation. This can be readily noticed by simply renaming the 'up' and 'down' single-particle states at one of the two stations.

This whole procedure goes against the grain of anyone who has followed the work of Bell-CHSH, because $\Delta$ contains the instrument settings of both experimental wings, which apparently spells some kind of non-locality. However, as explained above, the computation of the correlation uses judgements for equal and not-equal measurement outcomes i.e. judgements relative to the other experimental wing. Such judgements are naturally based on global facts as opposed to only local facts within the measurement stations. In fact, Bell-type constraints do not even rule out correlations of the form $E(\Delta) =-1+ 2 \left|\Delta (\mbox{mod} [-\pi,\pi))\right|/\pi$, which can be easily obtained in random games with macroscopic carriers, but are only purported to  rule out correlations depending on $\Delta$ in the form predicted by quantum mechanics.

\subsection{A perspective on Vorobev's cyclicities in terms of gauge symmetries}

\begin{figure}
\begin{center}
\includegraphics[width=10cm]{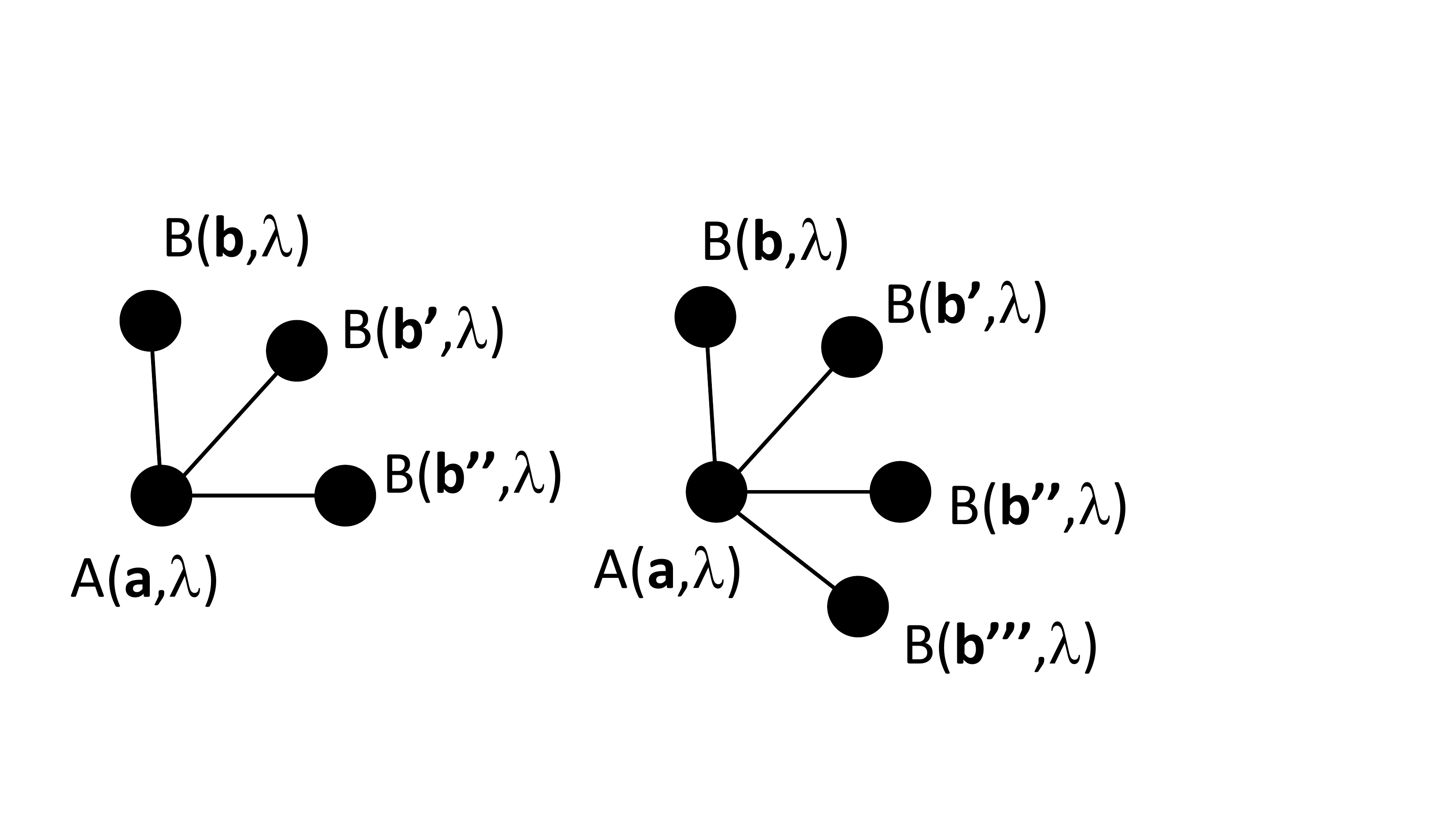}
\end{center}
\caption{Removed Vorob'ev cyclicity for Bell- and CHSH-type inequalities}
\label{fig:BerryBell}
\end{figure}

The authors of this present work have more recently proposed that relativity and gauge symmetry considerations permit and actually demand steps that remove the Vorob'ev cyclicity for a correct theoretical analysis of actual EPRB experiments \cite{Oaknin, Oaknin1, Oaknin2, HESS18}. These considerations follow the observation that the absolute direction of the polarizers or magnets in a Bell experiment is a redundant gauge variable, while the relative orientation of polarizers or magnets is the only true physical variable. A most important consequence of this observation is that we may and even must remove the cyclicity in the Bell-type inequalities as shown in Fig.\ref{fig:BerryBell}. Here we have fixed the instrument setting in one wing as a reference direction and just chosen the instrument settings in the other wing to obtain the correct Bell-CHSH angle-differences. As we shall show in what follows this is the only choice for which we can avoid with certainty assigning a double identity to any of the possible hidden events (which is the reason why it is usually said that some joint probabilities cannot be described on a single space).
As one can see, the cyclicity is removed and so is the constraint by Bell-CHSH inequalities and the equivalent inequalities given by Wigner's procedure. The probability distribution of the elements of reality emanating from the source is not changed by this procedure as it must not be. However, the evaluation of the relative outcomes by the measurement instruments may change and give different numbers for the 'equal' and 'not-equal' outcomes, depending on the relative instrument settings. Under certain circumstances that we shall now detail, this procedure is essential in order to make it possible to describe all the involved pair-wise random games on a single probability space.

We assume that the space of random events available at every repetition of the experiment form an unbiased sample within the whole space of events, so that we can consider for the sake of simplicity that the latter is always available . 

Thus, let $\left(\Omega, \Sigma, \mu\right)$ be a probability space, where $\Omega$ is a non-empty sample space, $\Sigma$ is the $\sigma$-algebra of all its subsets and $\mu$ is a (probability) measure defined on it, and let $\xi: \Omega \rightarrow \left[-1, +1\right] \subset \mathbb{R}$ be a random variable defined on it that takes values on the real interval $[-1, +1]$. 

Furthermore, let $\left\{{\cal F}\right\}_{\Delta \in Z}$ be a (continuous or discrete) group of isomorphic parameterizations of the probability space labelled over an additive group $Z$. 
That is, ${\cal F}_{\Delta} : \Omega \rightarrow \Omega$ are bijective applications from the sample space onto the sample space that preserve the probability measure:
\begin{equation}
\label{free_will}
(\forall \ S \subseteq \Omega) (\mu({\cal F}_{\Delta}[S])=\mu(S)).
\end{equation}
In particular, ${\cal F}_0 = {\mathbb I} : \Omega \rightarrow \Omega$ denotes the identity transformation.

Thus, two observers related by a relative 'displacement' $\Delta$ would describe the same random event, respectively, as $S \subseteq \Omega$ and ${\cal F}_{\Delta}[S] \subseteq \Omega$. Hence, the correlation between their descriptions of the random variable $\xi$ would be given by:
\begin{equation}
\mathbb{E}(\Delta) = \int_{\Omega} \ d\mu(w) \ \xi(w) \cdot \xi({\cal F}_{\Delta}(w)).    
\end{equation}

Bell-type inequalities constrain the two-parties correlations  $\left\{\mathbb{E}(\Delta_i)\right\}_{i=1,2,...,n}$ that can exist between parties for which, see Fig.\ref{fig:VorobCHSH},
\begin{equation}
\label{cyclicity}
\sum_{i=1,2,...,n} \Delta_i = 0,
\end{equation}
assuming that this cyclicity constraint requires that
\begin{equation}
\label{Bell}
{\cal F}_n \circ ... \ ... \circ {\cal F}_2 \circ {\cal F}_1 = {\cal F}_0.
\end{equation}
However, we notice that whenever gauge symmetries are involved a non-zero geometric (Berry) phase $\alpha \in Z$ may appear through some finite cyclic sequences [\ref{cyclicity}]:
\begin{equation}
\label{Berry}
{\cal F}_n \circ ... \ ... \circ {\cal F}_2 \circ {\cal F}_1 = {\cal F}_{\alpha} \neq {\cal F}_0.
\end{equation}

In other words, we consider the case in which cyclic sequences may be associated by a re-definition of the identity of symmetric events \cite{Oaknin,Oaknin1,Oaknin2,HESS18}, so that the parties cannot all be described within a single probability space. In such a case the same symmetry consideration can and must be used in order to remove the cyclicities, as shown in Fig.\ref{fig:BerryBell}. Obviously, this freedom cannot be allowed when all parties can test the random events at once, since it would imply that the events could have a ``double-identity" for at least one of the involved parties. On the other hand, this freedom must be considered in cases in which every random event can be tested only by a strict subset of the involved parties. In this case, the freedom [\ref{Berry}] is equivalent to stating that the identities of single parties cannot be properly defined, but only their relative 'displacements'. In physical terms we shall say that the identity of the parties is a gauge (non-physical) degree of freedom. This gauge freedom is tantamount to relaxing the cyclicity constraint [\ref{cyclicity}] and, therefore, it allows us to avoid the constraints that would appear otherwise. 

\subsubsection{Remark on the concept of contextuality}

 The notion of {\it contextuality}, frequently included in discussions about quantum foundations, actually corresponds in the case of Bell-type tests to different choices for the gauge-fixing condition in which the physical system is described. Hence, the {\it contextuality} of quantum phenomena may be removed by appropriately choosing a common gauge-fixing condition, namely, the common orientation of the measurement device in one of the two stations as shown in Fig.\ref{fig:BerryBell}.

\subsubsection{Remark on the relationship to experiments}

It is important to realize that the removal of the Vorob'ev cyclicity as discussed above is not necessarily associated with an actual geometrical rearrangement of the sources and measuring instruments involved in the EPRB experiments. This removal may be accomplished as a part of the theoretical description and analysis of the experiment by taking  advantage of the involved gauge symmetries and the fact that the polarization of each particle of every pair of entangled photons or electrons can be tested only along a single orientation. In fact, when a non-zero Berry phase [\ref{Berry}] appears through the considered cyclic arrangement it is a must to remove the cyclicity in order to describe all the involved pair-wise experiments together.

Thus, our analysis applies to both, experiments in which the detectors at the two experimental wings can be actually rotated \cite{Exper} and to the experiment of Giustina et al. \cite{Giustina}, which uses electro optical modulators (EOMs) that may actually be located at any place between source and detectors in both experimental wings.  The analysis applies as well to hypothetical experiments, not yet performed, with different sources emitting the entangled pairs and detectors arranged geometrically in a Vorob'ev cyclicity. Current optical fiber technology does appear to permit the construction of such experiments.

\section{Conclusion}

Bell-type inequalities for random games involving at least three functions defined on a single probability space have been known to mathematicians since the early works of Boole on probability theory \cite{Boole}, and found their more general formulation in the work of  Vorob'ev \cite{Vorob}. These inequalities constrain the correlations between pairs of random variables that can appear in this kind of games, and it is known that they are associated to certain cyclicities in the way how the variables are concatenated. The violation of the inequalities is understood to imply that the variables cannot be jointly defined on a single probability space.

In particular, the violation of this kind of inequalities by the correlations predicted by Quantum Mechanics has been understood as the ultimate  proof of the impossibility to describe quantum phenomena in terms of any underlying more fundamental theory based on the same fundamental physical principles derived from our macroscopic experience, and thus it represents the ultimate roadblock towards an intuitive interpretation of the quantum formalism.

In this paper we have shown, however, how subtle physically motivated considerations related to the gauge symmetries involved in the considered games may allow to remove the cyclicities and, hence, lift the constraints derived from the inequalities, paving the way to an explanation of  the quantum formalism within the framework of standard Einstein's causality and Kolmogorov's probability theory \cite{Oaknin,Oaknin1,Oaknin2,HESS18}.

As a byproduct of our symmetry argumentation we have presented an elegant way of obtaining the correlations predicted by quantum mechanics for the Bell experiment (\ref{july22n1}) from symmetry principles using standard tools of information theory.

Acknowledgement: We appreciate important comments by Hans De Raedt, Louis Marchildon and Andrei Khrennikov.


\end{document}